\documentclass[aps,prd,preprint,groupedaddress]{revtex4-1}
\usepackage[utf8]{inputenc}
\usepackage{physics}
\usepackage{appendix}
\usepackage{lipsum}
\usepackage{mathrsfs}
\usepackage[colorlinks=true,citecolor=blue,linkcolor=blue]{hyperref}
\usepackage[english]{babel}
\usepackage[T1]{fontenc}
\usepackage[Lenny]{fncychap}
\usepackage[left=2cm,right=2cm,top=2cm,bottom=2cm]{geometry}
\usepackage{amsmath,amssymb,amsthm,dsfont}
\usepackage{amsfonts}
\usepackage{graphicx}
\usepackage{subfig}
\usepackage{wrapfig}
\usepackage{slashed}
\usepackage{fancyhdr}
\selectfont
\usepackage{babel}

\begin{document}


\title{Influence of intense laser fields on measurable quantities in $W^{-}$-boson decay}


\author{M. Jakha,$^1$ S. Mouslih,$^{2,1}$  S. Taj,$^1$ and B. Manaut$^{1,}$}
\email[]{b.manaut@usms.ma}
\affiliation{ $^1$ Sultan Moulay Slimane University, Polydisciplinary Faculty,\\Research Team in Theoretical Physics and Materials (RTTPM), Beni Mellal, 23000, Morocco.\\$^2$ Sultan Moulay Slimane University, Faculty of Sciences and Techniques,\\ 
		Laboratory of Materials Physics (LMP),
		Beni Mellal, 23000, Morocco.}


\date{\today}

\begin{abstract}
In principle, this paper suggests powerful laser technology as a promising instrument that can be experimentally useful to control the lifetime and branching ratio for an unstable particle decay. In a recent paper [arXiv:2101.00224], we calculated theoretically the $W^{-}$-boson leptonic decay $(W^{-}\rightarrow \ell^{-} \bar{\nu}_{\ell})$ in the presence of a circularly polarized laser and we showed that the laser significantly contributed to the diminution of the leptonic decay rate. In this paper, as a continuation of the previous one, we mainly deal with the theoretical calculation of the $W^{-}$-boson hadronic decay $(W^{-}\rightarrow q \bar{q}')$ and we combine the analytical results obtained in both papers to examine the effect of an intense laser, in terms of its field strength and frequency, on the three measurable quantities in $W^{-}$-boson decay (total decay rate, lifetime and branching ratios). It was found that the laser has notably contributed to the reduction of the total decay rate leading to a longer lifetime. Most importantly, the two branching ratios (one for leptons and the other for hadrons) are affected (increased or decreased) by the presence of a strong external electromagnetic field. Combined together, these two complementary works may provide an in-depth and comprehensive study that would be useful for any experimental investigation in the future.
\end{abstract}

\maketitle
\section{Introduction}
It is undoubtedly that the laser has become, since its invention in the 1960s, one of the indispensable laboratory equipment used throughout the world \cite{bahk,yanovski}. The tremendous development that laser has made recently, whether in its intensities or its sources, has pushed both theoretical and experimental studies forward and has motivated physicists to carry out an increasing number of investigations to explore the interactions of high-intensity electromagnetic (EM) fields with matter \cite{salamin}. It is thanks to it that the study of the basic processes of quantum electrodynamics (QED) \cite{qed1,qed2,hartin,ritus1,ritus2,ehlotzky} and electroweak theory \cite{ritus3,becker} in the presence of a powerful laser field has been able to constantly attract the attention of researchers. In this context and as pointed out in the title, this research aims to study one of the fundamental processes of the electroweak theory, which is the decay of the boson $W^{-}$ in the presence of a strong EM field. There is no exaggeration in saying that the $W$ bosons and their neutral partner, the $Z$ boson, have been at the center of attention in high-energy physics for more than the past three decades \cite{rubbia}. Essentially, the discovery of these two bosons in 1983 by the UA1 \cite{ua1w,ua1z} and UA2 \cite{ua2w,ua2z} collaborations at CERN's $p\bar{p}$ collider provided direct confirmation of the unification of weak and EM interactions within a common framework called the electroweak theory predicted by physicists Sheldon Lee Glashow \cite{glashow}, Steven Weinberg \cite{weinberg} and Abdus Salam \cite{salam} in the late $1960$s. The properties of these particles, which are considered as mediators of the weak interaction, have been accurately measured in experiments at the Large Electron-Positron Collider (LEP) and the Stanford Linear Collider (SLC). The results of these experiments have made it possible to determine, among other things, the mass of the $W$-boson $M_{W}=80.379\pm0.012~\text{GeV}$ and its total decay rate $\Gamma_{W}=2.085\pm 0.042~\text{GeV}$ with a high degree of accuracy \cite{pdg2020}. In the Standard Model of electroweak interaction, the $W^{-}$-boson has two channels of decay. It can decay into a lepton $\ell^{-}$ and corresponding antineutrino $\bar{\nu}_{\ell}$ $(W^{-}\rightarrow \ell^{-} \bar{\nu}_{\ell})$ (leptonic channel) where $\ell=e,~\mu,~\tau$ or into a pair of quarks $(W^{-}\rightarrow q \bar{q}')$ (hadronic channel) where $q$ is $d$ or $s$ and $q'$ is the appropriate Cabibbo-Kobayashi-Maskawa (CKM) mixture of $u$ and $c$. Other hadronic decay channels are greatly suppressed by CKM off-diagonal matrix elements. As a first step, Mouslih \textit{et al.} studied recently the leptonic decay of the boson $W^{-}$ in the presence of a circularly polarized EM field and examined the effect of laser parameters on the leptonic decay rate \cite{mouslih1}. In a second step in this context, this paper comes to highlight mainly the hadronic decay of the boson $W^{-}$. We will combine the results obtained with the previous ones in order to be able to study the effect of the laser on the total decay rate and the lifetime as well as on the branching ratios (BRs). Therefore, this present work is in fact an extension and a continuation of the first one. The results obtained there are required to extract those presented here. The reader can refer to our previous article \cite{mouslih1} for more references and documentation on this topic. The rest of the paper is planned as follows. In Sec.~\ref{sec:theory}, we will try to formulate, in detail, the theoretical expressions of the hadronic decay rate, lifetime and branching ratios. Then, in Sec.~\ref{sec:results}, we will present and discuss the results obtained. The last Sec.~\ref{sec:conclusion} is devoted to summarizing our conclusions. We only mention here that we used the well-known natural units $c=1=\hbar$ during our calculations.  
\section{Outline of the theory}\label{sec:theory}
\subsection{Hadronic decay rate in the presence of a laser field with circular polarization}
\begin{figure}[hbtp]
\centering
\includegraphics[scale=0.6]{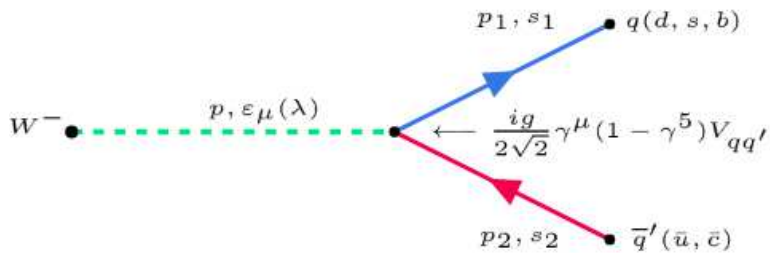}
\caption{Lowest order Feynman diagram of the hadronic  $W^{-}$-boson decay.}\label{diagram}
\end{figure}

In the framework of electroweak theory, we consider the hadronic decay of the negatively charged $W^{-}$-boson into a pair of quarks $(W^{-}\rightarrow q \bar{q}')$, where $q$ or $q'$ represent one of the quarks $u$, $d$, $c$, $s$, or $b$ except $t$ since top quark is heavier than the $W$ boson. The corresponding Feynman diagram is shown in Fig.~\ref{diagram}. 
We take the laser field as a circularly polarized monochromatic EM field, which is classically described by the following four-potential:
\begin{align}\label{potential}
A^{\mu}(\phi)=a^{\mu}_{1}\cos(\phi)+a^{\mu}_{2}\sin(\phi),
\end{align}
depending on a single variable, $\phi=(k.x)$, the phase of the laser field. $k=(\omega,\textbf{k})$ is the wave four-vector $(k^{2}=0)$ and $\omega$ is the laser frequency. The  four-amplitudes $a^{\mu}_{1}=|\textbf{a}|(0,1,0,0)$ and $a^{\mu}_{2}=|\textbf{a}|(0,0,1,0)$ are equal in magnitude and orthogonal, which implies $(a_{1}.a_{2})=0$ and $a_{1}^{2}=a_{2}^{2}=a^{2}=-|\textbf{a}|^{2}=-(\mathcal{E}_{0}/\omega)^{2}$ where $\mathcal{E}_{0}$ is the amplitude of the electric field. We suppose that the four-potential satisfies the Lorentz gauge condition, $k_{\mu}A^{\mu}=0$, which means $(k.a_{1})=(k.a_{2})=0$, indicating that the wave vector $\textbf{k}$ is chosen to be along the $z$-axis. \\
The decay process of the $W^{-}$-boson  in the field of a circularly polarized EM plane wave is a weak interaction process; it can be described by the lowest Feynman diagrams. Therefore, the lowest-order $S$-matrix element for the laser-assisted hadronic  $W^{-}$ decay reads \cite{greiner}
\begin{equation}\label{smatrix}
S_{fi}(W^{-}\rightarrow q \bar{q}')=\dfrac{i\textsl{g}~ V_{qq'}}{2\sqrt{2}}\int d^{4}x\overline{\psi}_{q}(x)\gamma^{\mu}(1-\gamma^{5})\psi_{\bar{q}'}(x)W^{-}_{\mu}(x),
\end{equation}
where $\textsl{g}$ is the electroweak coupling constant and $V_{qq'}$ is the Cabibbo-Kobayashi-Maskawa (CKM) quark-mixing matrix corresponding element. In order to take into account the interaction of the incoming electrically charged $W^{-}$-boson (1-spin particle) with the EM field, we will describe it by the following wave function \cite{kurilin1999,obukhov2}:
\begin{equation}\label{wwave function}
W^{-}_{\mu}(x)=\bigg[\textsl{g}_{\mu\nu}-\dfrac{e}{(k.p)}\big(k_{\mu}A_{\nu}-k_{\nu}A_{\mu}\big)-\frac{e^{2}}{2(k.p)^{2}}A^{2}k_{\mu}k_{\nu}\bigg]\frac{\varepsilon^{\nu}(p,\lambda)}{\sqrt{2p_{0}V}}\times e^{iS(q,x)},
\end{equation}
where $\textsl{g}_{\mu\nu}=\text{diag}(1,-1,-1,-1)$ is the metric tensor of Minkowski space, $\varepsilon^{\nu}(p,\lambda)$ is the $W^{-}$-boson polarization four-vector such that the summation over all three directions of polarization $\lambda$ yields $\sum_{\lambda=1}^{3}\varepsilon_{\mu}(p,\lambda)\varepsilon_{\nu}^{*}(p,\lambda)=-\textsl{g}_{\mu\nu}+p_{\mu}p_{\nu}/M_{W}^{2}$, and
\begin{equation}
S(q,x)=-q.x+\dfrac{e(a_{1}.p)}{k.p}\sin(\phi)-\dfrac{e(a_{2}.p)}{k.p}\cos(\phi),
\end{equation}
with the dressed four-momentum $q=(Q,\textbf{q})$ and the effective mass $M_{W}^{*}$ that the boson $W^{-}$ acquires inside the EM field are, respectively, such that
\begin{align}
q=p-\frac{e^{2}a^{2}}{2(k.p)}k,~~~~ M^{*}_{W}=\sqrt{M_{W}^{2}-e^{2}a^{2}}, 
\end{align}
where $M_{W}$ is the rest mass of the $W^{-}$-boson. The outgoing  quarks are described by the relativistic Dirac-Volkov functions normalized to the volume $V$ \cite{volkov}:
\begin{equation}\label{hwave function}
\begin{split}
&\psi_{q}(x)=\bigg[1+\dfrac{\eta e\slashed{k}\slashed{A}}{2(k.p_{1})}\bigg]\frac{u(p_{1},s_{1})}{\sqrt{2Q_{1}V}}\times e^{iS(q_{1},x)},\\
&\psi_{\bar{q}'}(x)=\bigg[1-\dfrac{\eta' e\slashed{k}\slashed{A}}{2(k.p_{2})}\bigg]\frac{v(p_{2},s_{2})}{\sqrt{2Q_{2}V}}\times e^{-iS(q_{2},x)},
\end{split}
\end{equation}
where the two factors $\eta=1/3$ and $\eta'=-2/3$ are, respectively, due to the fractional charge of down and up quarks; regarding the sign one should note that $e=-|e|<0$ is the charge of the electron, and
\begin{equation}
\begin{split}
&S(q_{1},x)=-q_{1}.x-\dfrac{\eta e(a_{1}.p_{1})}{k.p_{1}}\sin(\phi)+\dfrac{\eta e(a_{2}.p_{1})}{k.p_{1}}\cos(\phi),\\
&S(q_{2},x)=-q_{2}.x-\dfrac{\eta' e(a_{1}.p_{1})}{k.p_{2}}\sin(\phi)+\dfrac{\eta' e(a_{2}.p_{2})}{k.p_{2}}\cos(\phi).
\end{split}
\end{equation}
$u(p_{1},s_{1})$ and $v(p_{2},s_{2})$ represent the Dirac bispinors for the outgoing quarks  with momentum $p_{i=1,2}$ and spin $s_{i=1,2}$ satisfying
\begin{align}
\begin{split}
&\sum_{s_{1}}u(p_{1},s_{1})\overline{u}(p_{1},s_{1})=\slashed{p}_{1}+m_{q},\\
&\sum_{s_{2}}v(p_{2},s_{2})\overline{v}(p_{2},s_{2})=\slashed{p}_{2}-m_{q'},
\end{split}
\end{align}
where $m_{q}$ and $m_{q'}$ are, respectively, the rest masses of the quark $q$ and antiquark $q'$. The four-momentum $q_{i}(i=1,2)  = (Q_{i} , \textbf{q}_{i} )$ is the quasi-momentum that the quark and antiquark acquire in the presence of the EM field
\begin{equation}
\begin{split}
q_{1}=&p_{1}-\frac{(\eta e)^{2}a^{2}}{2(k.p_{1})}k,\\
q_{2}=&p_{2}-\frac{(\eta' e)^{2}a^{2}}{2(k.p_{2})}k.
\end{split}
\end{equation}
To obtain analytical results, we proceed as follows: we insert Eqs.~(\ref{wwave function}) and (\ref{hwave function}) into Eq.~(\ref{smatrix}) and after some manipulations, we find the $S$-matrix to be written as
\begin{equation}\label{ssmatrix}
\begin{split}
S_{fi}(W^{-}\rightarrow q \bar{q}')=&\dfrac{i\textsl{g} ~ V_{qq'}}{2\sqrt{2}\sqrt{8 Q_{2}Q_{1}p_{0}V^{3}}}\int d^{4}x\overline{u}(p_{1},s_{1})\Big\lbrace\Big[1+C(p_{1})\slashed{A}\slashed{k}\Big]\gamma^{\mu}(1-\gamma^{5})\\&\times\Big[1-C(p_{2})\slashed{k}\slashed{A}\Big]\Big[\textsl{g}_{\mu\nu}-C(p)\big(k_{\mu}A_{\nu}-k_{\nu}A_{\mu}\big)-\frac{C(p)^{2}}{2}a^{2}k_{\mu}k_{\nu}\Big]\Big\rbrace \\
&\times v(p_{2},s_{2})\varepsilon^{\nu}(p,\lambda) e^{i(S(q,x)-S(q_{1},x)-S(q_{2},x))},
\end{split}
\end{equation}
where 
\begin{align*}
C(p_{1})=\dfrac{\eta e}{2(k.p_{1})},~~~~C(p_{2})=\dfrac{\eta' e}{2(k.p_{2})},~~~~ C(p)=\dfrac{e}{(k.p)}.
\end{align*}
Now, let us transform the exponential term by introducing the following parameter:
\begin{equation}\label{argument}
\begin{split}
&z=\sqrt{\alpha_{1}^{2}+\alpha_{2}^{2}}\quad\text{with}\quad\alpha_{1}=-e\bigg(\dfrac{a_{1}.p}{k.p}+\dfrac{\eta (a_{1}.p_{1})}{k.p_{1}}+\dfrac{\eta' (a_{1}.p_{2})}{k.p_{2}}\bigg),\\
&\quad\alpha_{2}=-e\bigg(\dfrac{a_{2}.p}{k.p}+\dfrac{\eta(a_{2}.p_{1})}{k.p_{1}}+\dfrac{\eta' (a_{2}.p_{2})}{k.p_{2}}\bigg),
\end{split}
\end{equation}
this yields
\begin{equation}
e^{i(S(q,x)-S(q_{1},x)-S(q_{2},x))}= e^{i(q_{1}+q_{2}-q).x}e^{-iz\sin(\phi-\phi_{0})},
\end{equation}
with $\phi_{0}=\arctan(\alpha_{2}/\alpha_{1})$. Therefore, the $S$-matrix element becomes
\begin{equation}\label{smatrix2}
\begin{split}
S_{fi}(W^{-}\rightarrow q \bar{q}')=&\dfrac{i\textsl{g}~V_{qq'}}{2\sqrt{2}\sqrt{8Q_{2}Q_{1}p_{0}V^{3}}}\int d^{4}x\overline{u}(p_{1},s_{1})\big[\Delta_{\nu}^{0}+ \Delta_{\nu}^{1}\cos(\phi)+\Delta_{\nu}^{2}\sin(\phi)\big]\\&\times v(p_{2},s_{2})\varepsilon^{\nu}(p,\lambda) e^{i(q_{1}+q_{2}-q).x}e^{-iz\sin(\phi-\phi_{0})},
\end{split}
\end{equation}
where the three quantities $\Delta_{\nu}^{0}$, $\Delta_{\nu}^{1}$ and $\Delta_{\nu}^{2}$ are expressed as follows:
\begin{equation}
\begin{split}
\Delta_{\nu}^{0}=&\gamma^{\mu}\big(1-\gamma^{5}\big)\big(\textsl{g}_{\mu\nu}-\frac{C(p)^{2}}{2}a^{2}k_{\mu}k_{\nu}\big)+2k_{\nu}a^{2}\slashed{k}\big(1-\gamma^{5}\big),\\
\Delta_{\nu}^{1}=&-C(p_{2})\gamma^{\mu}\big(1-\gamma^{5}\big)\slashed{k}\slashed{a}_{1}\big(\textsl{g}_{\mu\nu}-\frac{C(p)^{2}}{2}a^{2}k_{\mu}k_{\nu}\big)+C(p_{1})\slashed{a}_{1}\slashed{k}\gamma^{\mu}\big(1-\gamma^{5}\big)\\
&\times(\textsl{g}_{\mu\nu}-\frac{C(p)^{2}}{2}a^{2}k_{\mu}k_{\nu}\big)-C(p)\gamma^{\mu}\big(1-\gamma^{5}\big)\big(k_{\mu}a_{1\nu}-k_{\nu}a_{1\mu}\big),\\
\Delta_{\nu}^{2}=&-C(p_{2})\gamma^{\mu}\big(1-\gamma^{5}\big)\slashed{k}\slashed{a}_{2}\big(\textsl{g}_{\mu\nu}-\frac{C(p)^{2}}{2}a^{2}k_{\mu}k_{\nu}\big)+C(p_{1})\slashed{a}_{2}\slashed{k}\gamma^{\mu}\big(1-\gamma^{5}\big)\\
&\times(\textsl{g}_{\mu\nu}-\frac{C(p)^{2}}{2}a^{2}k_{\mu}k_{\nu}\big)-C(p)\gamma^{\mu}\big(1-\gamma^{5}\big)\big(k_{\mu}a_{2\nu}-k_{\nu}a_{2\mu}\big).
\end{split}
\end{equation}
It is worth noting here that one must be careful, throughout the calculation,  with the positions of the indices $(\mu$ and $\nu)$ that must be respected in order to be contracted.\\
The linear combination of the three different quantities in Eq.~(\ref{smatrix2}) can be transformed by the well-known identities involving ordinary Bessel functions $J_{s }(z)$ \cite{landau}:
\begin{align}
\begin{bmatrix}
1\\
\cos(\phi)\\
\sin(\phi)
\end{bmatrix}\times e^{-iz\sin(\phi-\phi_{0})}=\sum_{s=-\infty}^{+\infty}\begin{bmatrix}
B_{s}(z)\\
B_{1s}(z)\\
B_{2s}(z)
\end{bmatrix}e^{-is\phi},
\end{align}
where
\begin{align}
\begin{split}
\begin{bmatrix}
B_{s}(z)\\
B_{1s}(z)\\
B_{2s}(z) \end{bmatrix}=\begin{bmatrix}J_{s}(z)e^{is\phi_{0}}\\
\big(J_{s+1}(z)e^{i(s+1)\phi_{0}}+J_{s-1}(z)e^{i(s-1)\phi_{0}}\big)/2\\
\big(J_{s+1}(z)e^{i(s+1)\phi_{0}}-J_{s-1}(z)e^{i(s-1)\phi_{0}}\big)/2i
 \end{bmatrix},
\end{split}
\end{align}
where $z$ is the argument of the Bessel functions defined in Eq.~(\ref{argument}) and $s$, their order, is interpreted as the number of exchanged photons. Using these transformations in Eq.~(\ref{smatrix2}) and integrating over $d^{4}x$, the matrix element $S_{fi}$ becomes
\begin{equation}\label{smatrix3}
S_{fi}(W^{-}\rightarrow q \bar{q}')=\dfrac{i\textsl{g}~V_{qq'}}{2\sqrt{2}\sqrt{8Q_{2}Q_{1}p_{0}V^{3}}}\sum_{s=-\infty}^{\infty}\mathcal{M}^{s}_{fi}(2\pi)^{4}\delta^{4}(q_{1}+q_{2}-q-sk),
\end{equation}
where the quantity $\mathcal{M}^{s}_{fi}$ is defined by
\begin{align}
\mathcal{M}^{s}_{fi}=\bar{u}(p_{1},s_{1})\Lambda_{\nu}^{s}v(p_{2},s_{2})\varepsilon^{\nu}(p,\lambda),
\end{align}
where
\begin{align}
\Lambda_{\nu}^{s}=\Delta_{\nu}^{0}B_{s}(z)+ \Delta_{\nu}^{1}B_{1s}(z)+\Delta_{\nu}^{2}B_{2s}(z).
\end{align}
The decay rate of the $W^{-}$-boson is obtained by multiplying the squared $S$-matrix element by the density of final states, summing over spins of quarks and antiquarks, averaging over the polarization of the boson $W^{-}$, and finally dividing by the time $T$. We get
\begin{align}\label{summed}
\Gamma(W^{-}\rightarrow q \bar{q}')=\sum_{s=-\infty}^{+\infty}\Gamma^{s}(W^{-}\rightarrow q \bar{q}'),
\end{align}
where the photon-number-resolved decay rate $\Gamma^{s}(W^{-}\rightarrow q \bar{q}')$ is defined by
 \begin{align}
\Gamma^{s}(W^{-}\rightarrow q \bar{q}')=\dfrac{\textsl{g}^{2} |V_{qq'}|^{2}N_{c}}{64 p_{0}}\int\dfrac{d^{3}q_{1}}{(2\pi)^{3}Q_{1}}\int\dfrac{d^{3}q_{2}}{(2\pi)^{3}Q_{2}}(2\pi)^{4}\delta^{4}(q_{1}+q_{2}-q-sk)|\overline{\mathcal{M}^{s}_{fi}}|^{2},
\end{align}
where $\textit{N}_{c}=3$ is the number of colors, and
\begin{align}\label{sums}
|\overline{\mathcal{M}^{s}_{fi}}|^{2}=\frac{1}{3}\sum_{\lambda}\sum_{s_{1},s_{2}}|\mathcal{M}^{s}_{fi}|^{2}=\frac{1}{3}\sum_{\lambda}\sum_{s_{1},s_{2}}|\bar{u}(p_{1},s_{1})\Lambda_{\nu}^{s}v(p_{2},s_{2})\varepsilon^{\nu}(p,\lambda)|^{2}.
\end{align}
Performing the integration over $d^{3}q_{2}$ and using $\delta^{4}(q_{1}+q_{2}-q-sk)=\delta^{3}(\textbf{q}_{1}+\textbf{q}_{2}-\textbf{q}-s\textbf{k})\delta(Q_{1}+Q_{2}-Q-s\omega)$, the photon-number-resolved decay rate $\Gamma^{s}$ becomes
\begin{align}
\Gamma^{s}(W^{-}\rightarrow q \bar{q}')=\dfrac{\textsl{g}^{2} |V_{qq'}|^{2}N_{c}}{64(2\pi)^{2}p_{0}}\int\dfrac{d^{3}q_{1}}{Q_{1}Q_{2}}\delta(Q_{1}+Q_{2}-Q-s\omega)|\overline{\mathcal{M}^{s}_{fi}}|^{2},
\end{align}
with $\textbf{q}_{1}+\textbf{q}_{2}-\textbf{q}-s\textbf{k}=0$. We choose the $W^{-}$-boson rest frame in which $Q=M^{*}_{W}$ and $\textbf{q}=0$, then $\textbf{q}_{2}=s\textbf{k}-\textbf{q}_{1}$. Hence, using $ d^{3}q_{1}=|\textbf{q}_{1}|^{2}d|\textbf{q}_{1}|d\Omega_{q}$, we obtain
\begin{align}
\begin{split}
\Gamma^{s}(W^{-}\rightarrow q \bar{q}')=&\dfrac{\textsl{g}^{2}|V_{qq'}|^{2}N_{c}}{64(2\pi)^{2}p_{0}}\int\dfrac{|\textbf{q}_{1}|^{2}d|\textbf{q}_{1}|d\Omega_{q}}{Q_{1}Q_{2}}\delta\Big(\sqrt{|\textbf{q}_{1}|^{2}+m_{q}^{*2}}\\&+\sqrt{(s\omega)^{2}+|\textbf{q}_{1}|^{2}-2s\omega|\textbf{q}_{1}|\cos(\theta)+m_{q'}^{*2}}-M^{*}_{W}-s\omega\Big)|\overline{\mathcal{M}^{s}_{fi}}|^{2},
\end{split}
\end{align}
where $m_{q}^{*}$ and $m_{q'}^{*}$ are, respectively, the effective masses of quarks $q$ and $q'$ given by
\begin{equation}\label{qeffmass}
\begin{split}
m_{q}^{*}&=\sqrt{m_{q}^{2}+(\eta e)^{2}\mathcal{E}_{0}^{2}/\omega^{2}},\\
m_{q'}^{*}&=\sqrt{m_{q'}^{2}+(\eta' e)^{2}\mathcal{E}_{0}^{2}/\omega^{2}}.
\end{split}
\end{equation}
Accordingly, the effective mass seems to be approximately linearly proportional to the field strength $\mathcal{E}_{0}$.\\
The remaining integral over $d|\textbf{q}_{1}|$ can be solved by using the familiar formula \cite{greiner}:
\begin{align}
\int dxf(x)\delta(g(x))=\dfrac{f(x)}{|g'(x)|}\bigg|_{g(x)=0}.
\end{align}
Thus we get
\begin{align}\label{sdecayrate}
\begin{split}
\Gamma^{s}(W^{-}\rightarrow q \bar{q}')=&\dfrac{\textsl{g}^{2}|V_{qq'}|^{2}N_{c}}{64(2\pi)^{2}p_{0}}\int\dfrac{|\textbf{q}_{1}|^{2}d\Omega_{q}}{Q_{1}Q_{2}}\dfrac{|\overline{\mathcal{M}^{s}_{fi}}|^{2}}{|g'(|\textbf{q}_{1}|)|}\bigg|_{g(|\textbf{q}_{1}|)=0},
\end{split}
\end{align}
where 
\begin{align}
g'(|\textbf{q}_{1}|)=\dfrac{|\textbf{q}_{1}|}{\sqrt{|\textbf{q}_{1}|^{2}+m_{q}^{*2}}}+\dfrac{|\textbf{q}_{1}|-s\omega\cos(\theta)}{\sqrt{(s\omega)^{2}+|\textbf{q}_{1}|^{2}-2s\omega|\textbf{q}_{1}|\cos(\theta)+m_{q'}^{*2}}}.
\end{align}
The sums over spins in the term $|\overline{\mathcal{M}^{s}_{fi}}|^{2}$ (Eq.~(\ref{sums})) can be reduced to the calculation of traces as follows:
\begin{align}
|\overline{\mathcal{M}^{s}_{fi}}|^{2}=\frac{1}{3}\bigg(-\textsl{g}^{\mu\nu}+\frac{p^{\mu}p^{\nu}}{M_{W}^{2}}\bigg)\text{Tr}\big[(\slashed{p}_{1}+m_{q})\Lambda_{\nu}^{s}(\slashed{p}_{2}-m_{q'})\overline{\Lambda}_{\mu}^{s}\big],
\end{align}
where
\begin{align}
\begin{split}
\overline{\Lambda}_{\mu}^{s}&=\gamma^{0}\Lambda_{\mu}^{s\dagger}\gamma^{0},\\
&=\overline{\Delta}_{\mu}^{0}B^{*}_{s}(z)+ \overline{\Delta}_{\mu}^{1}B^{*}_{1s}(z)+\overline{\Delta}_{\mu}^{2}B^{*}_{2s}(z),
\end{split}
\end{align}
and
\begin{equation}
\begin{split}
\overline{\Delta}_{\mu}^{0}=&\gamma^{0}\Delta_{0}^{\dagger}\gamma^{0}=\gamma^{\nu}\big(1-\gamma^{5}\big)\big(\textsl{g}_{\mu\nu}-\frac{C(p)^{2}}{2}a^{2}k_{\mu}k_{\nu}\big)+2k_{\mu}a^{2}\slashed{k}\big(1-\gamma^{5}\big),\\
\overline{\Delta}_{\mu}^{1}=&\gamma^{0}\Delta_{1}^{\dagger}\gamma^{0}=-C(p_{2})\slashed{a}_{1}\slashed{k}\gamma^{\nu}\big(1-\gamma^{5}\big)\big(\textsl{g}_{\mu\nu}-\frac{C(p)^{2}}{2}a^{2}k_{\mu}k_{\nu}\big)+C(p_{1})\gamma^{\nu}\big(1-\gamma^{5}\big)\slashed{k}\slashed{a}_{1}\\
&\times(\textsl{g}_{\mu\nu}-\frac{C(p)^{2}}{2}a^{2}k_{\mu}k_{\nu}\big)-C(p)\gamma^{\nu}\big(1-\gamma^{5}\big)\big(k_{\nu}a_{1\mu}-k_{\mu}a_{1\nu}\big),\\
\overline{\Delta}_{\mu}^{2}=&\gamma^{0}\Delta_{2}^{\dagger}\gamma^{0}=-C(p_{2})\slashed{a}_{2}\slashed{k}\gamma^{\nu}\big(1-\gamma^{5}\big)\big(\textsl{g}_{\mu\nu}-\frac{C(p)^{2}}{2}a^{2}k_{\mu}k_{\nu}\big)+C(p_{1})\gamma^{\nu}\big(1-\gamma^{5}\big)\slashed{k}\slashed{a}_{2}\\
&\times(\textsl{g}_{\mu\nu}-\frac{C(p)^{2}}{2}a^{2}k_{\mu}k_{\nu}\big)-C(p)\gamma^{\nu}\big(1-\gamma^{5}\big)\big(k_{\nu}a_{2\mu}-k_{\mu}a_{2\nu}\big).\\
\end{split}
\end{equation}
The trace calculation is performed with the help of FEYNCALC \cite{feyncalc1,feyncalc2,feyncalc3}. Refer to the Appendix for the detailed and explicit expression of $|\overline{\mathcal{M}^{s}_{fi}}|^{2}$.
\subsection{Lifetime and branching ratios}
In the previous section, we have demonstrated the theoretical expression of the hadronic decay rate in the presence of a circularly polarized EM field (Eq.~(\ref{sdecayrate})). The well-known and very important quantity that comes directly after the decay rate is the lifetime which is the inverse of the total decay rate. The latter is the sum of the leptonic and hadronic decay rates of the $W^{-}$-boson. Taking into account the analytical results obtained previously in the study of the $W^{-}$-boson leptonic decay $(W^{-}\rightarrow \ell^{-}\bar{\nu}_{\ell})$ \cite{mouslih1} and combining them with the current results of hadronic decay, we can obtain the expression of the lifetime in the presence of an EM field as follows:
\begin{equation}
\tau_{W}=\frac{1}{\Gamma_{W}^{\text{tot}}},
\end{equation}
where 
\begin{equation}
\Gamma_{W}^{\text{tot}}=\Gamma(W^{-}\rightarrow \text{leptons})+\Gamma(W^{-}\rightarrow \text{hadrons}).
\end{equation}
$\Gamma(W^{-}\rightarrow \text{leptons})$ is the total leptonic decay rate given by
\begin{equation}\label{leptrate}
\Gamma(W^{-}\rightarrow \text{leptons})=\Gamma(W^{-}\rightarrow e^{-}\bar{\nu}_{e})+\Gamma(W^{-}\rightarrow \mu^{-}\bar{\nu}_{\mu})+\Gamma(W^{-}\rightarrow \tau^{-}\bar{\nu}_{\tau}).
\end{equation}
For the total hadronic decay rate, we consider the two dominant hadronic channels
\begin{equation}\label{hadrate}
\Gamma(W^{-}\rightarrow \text{hadrons})=\Gamma(W^{-}\rightarrow \bar{u}d)+\Gamma(W^{-}\rightarrow \bar{c}s).
\end{equation}
After giving the definition of the lifetime, let us now introduce another very interesting and experimentally measurable quantity. This is the branching ratio (BR) defined as the ratio between each partial decay rate and the total decay rate. In particle physics, it indicates the probability that a particle will follow a specific decay channel among all possible decay ones. Thus, the sum of the BRs is equal to $1$ (or $100\%$). In our case, we define the BRs for the leptonic and hadronic decay channels as follows:
\begin{equation}\label{brhadrons}
\text{BR}(W^{-}\rightarrow \text{hadrons})=\frac{\Gamma(W^{-}\rightarrow \text{hadrons})}{\Gamma_{W}^{\text{tot}}}.
\end{equation}
\begin{equation}\label{brleptons}
\text{BR}(W^{-}\rightarrow \text{leptons})=\frac{\Gamma(W^{-}\rightarrow \text{leptons})}{\Gamma_{W}^{\text{tot}}},
\end{equation}
Their experimental values in the absence of the laser field are \cite{pdg2020}
\begin{equation}
\begin{split}
\text{BR}(W^{-}\rightarrow \text{hadrons})&=(67.41\pm 0.27)\%,\\
\text{BR}(W^{-}\rightarrow \text{leptons})&=(32.58\pm 0.16)\%.
\end{split}
\end{equation}
\subsection{Comparison with leptonic decay}
In this section, we would like to make a comparison between the leptonic and hadronic decays of the $W^{-}$-boson and give some distinctions between them. The process of leptonic decay $(W^{-}\rightarrow \ell^{-}\bar{\nu}_{\ell})$ contains, in its final state, an electrically neutral particle (antineutrino $\bar{\nu}_{\ell}$), and therefore its interaction with the EM field is not taken into account. Therefore, when applying the EM field, we only have to dress two particles, the incoming boson $W^{-}$ and the final charged lepton $\ell^{-}$. But, it is completely different in the case of hadronic decay $(W^{-}\rightarrow q \bar{q}')$ where all the particles involved are electrically charged and thus all of them will be dressed up and described theoretically by Volkov's functions. The calculation in the latter case is much more complicated and tedious compared to the first case. In the absence of the laser field and in the case of neglected fermion masses, it is sufficient to consider the color factor $N_{c}$ and the corresponding element in the CKM mixing matrix $V_{qq'}$ in order to approximate the hadronic decay to the leptonic one, for example
\begin{equation}
\Gamma(W^{-}\rightarrow \bar{u}d)\simeq 3|V_{ud}|^{2}\Gamma(W^{-}\rightarrow e^{-}\bar{\nu}_{e}),
\end{equation}
where the partial decay rate $\Gamma(W^{-}\rightarrow e^{-}\bar{\nu}_{e})$ is given, in the approximation of neglected masses, by \cite{greiner}
\begin{equation}
\Gamma(W^{-}\rightarrow e^{-}\bar{\nu}_{e})\simeq\frac{G_{F}M_{W}^{3}}{6\pi\sqrt{2}},
\end{equation}
where $G_{F}=(1.166~37\pm0.000~02)\times10^{-11}~\text{MeV}^{-2}$ is the Fermi coupling constant. Using the unitarity of the CKM matrix, the total hadronic decay rate is given by 
\begin{equation}
\Gamma(W^{-}\rightarrow \text{hadrons})=\Gamma(W^{-}\rightarrow \bar{u}d)+\Gamma(W^{-}\rightarrow \bar{c}s)\simeq 6\Gamma(W^{-}\rightarrow e^{-}\bar{\nu}_{e}),
\end{equation}
where we have considered the CKM matrix as identity matrix $(|V_{ud}|\simeq |V_{cs}|\simeq 1)$. The total decay rate of the $W^{-}$ is given by (sum over the three leptonic decays and the two hadronic decays)
\begin{equation}
\Gamma_{W}^{\text{tot}}\simeq 9\Gamma(W^{-}\rightarrow e^{-}\bar{\nu}_{e}).
\end{equation}
One must be careful not to apply this fact in the case where the EM field is present.
\section{Results and Discussion}\label{sec:results}
\begin{figure}[hbtp]
\centering
\includegraphics[scale=0.8]{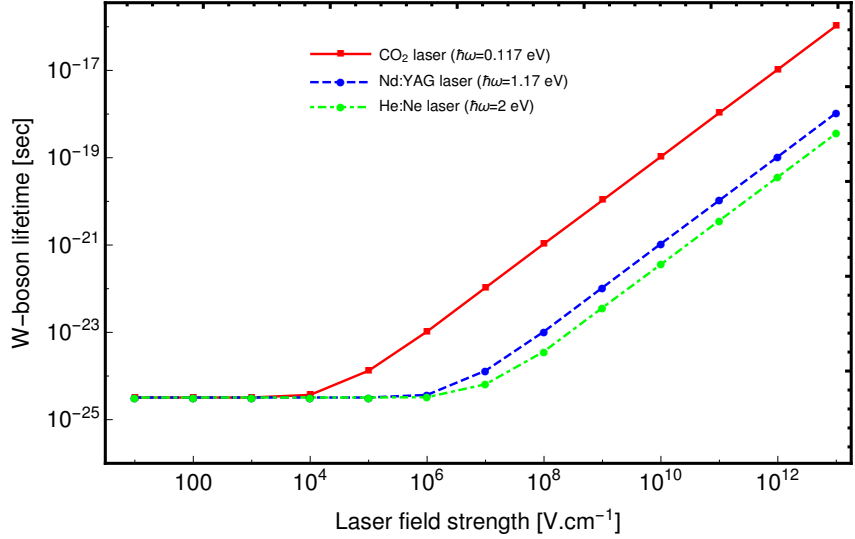}
\caption{The variations of laser-assisted $W$-boson lifetime as a function of the laser field strength for a CO$_{2}$ laser ($\hbar\omega=0.117~\text{eV}$), a Nd:YAG laser
($\hbar\omega=1.17~\text{eV}$) and a He:Ne laser ($\hbar\omega=2~\text{eV}$). The number of exchanged photons is taken as $-20\leq s\leq +20$.}\label{fig2}
\end{figure}

\begin{table}
\caption{\label{tab1}Numerical values of the ratio $R_{\text{with/without}}$, given in Eq.~(\ref{ratio}), as a function of the laser field strength $\mathcal{E}_{0}$ for three different frequencies (CO$_{2}$ laser: $\hbar\omega=0.117~\text{eV}$, Nd:YAG laser: $\hbar\omega=1.17~\text{eV}$ and He:Ne laser: $\hbar\omega=2~\text{eV}$). We have summed, in the presence of the laser, over the number of photons $-20\leq s\leq +20$.}
\begin{ruledtabular}
\begin{tabular}{cccc} 
 &\multicolumn{3}{c}{$R_{\text{with/without}}$}\\  \cline{2-4}
$\mathcal{E}_{0}$ $[\text{V/cm}]$ & $\hbar\omega=0.117~\text{eV}$  & $\hbar\omega=1.17~\text{eV}$ & $\hbar\omega=2~\text{eV}$ \\ \hline
 $10$       & $0.999963$        &  $0.999964$  &  $0.999964$ \\
 $10^{2}$   & $0.999963$       &   $0.999964$   &  $0.999964$\\
 $10^{3}$   & $0.999942$       &   $0.999963$   &  $0.999964$\\
 $10^{4}$   & $0.8713$       &    $0.999963$   & $0.999963$\\
 $10^{5}$   & $0.243064$       &   $0.999942$   & $0.999963$\\
 $10^{6}$   & $0.0307884$       &   $0.8713$  & $0.983108$\\
 $10^{7}$   &    $0.00303762$    & $0.243064$  & $0.497739$\\
 $10^{8}$   & $0.000300683$     &  $0.0307884$  & $0.0879964$\\
 $10^{9}$   & $0.0000304507$    & $0.00303762$ & $0.00876256$\\
 $10^{10}$  & $3.04641\times 10^{-6}$   & $0.000300683$ & $0.000892935$\\
 $10^{11}$  & $3.01282\times 10^{-7}$ & $0.000030451$ &$0.0000884333$\\
 $10^{12}$  & $3.06297\times 10^{-8}$ & $3.04709\times 10^{-6}$ & $8.91838\times10^{-6}$\\
 $10^{13}$  & $3.05141\times  10^{-9}$ & $3.02022\times 10^{-7}$&$8.85819\times10^{-7}$
\end{tabular}
\end{ruledtabular}
\end{table}
This paper investigates the effect of a strong and circularly polarized EM field on experimentally measurable quantities in the decay process of the boson $W^{-}$. During this section, we will try to present and analyze the numerical results obtained and discuss their physical interpretation. However, in any decay process, we have three very important quantities that are experimentally measurable. First, the total decay rate (i.e, the probability per unit time that the particle will decay) measured via the so-called Breit-Wigner distribution. Secondly, the lifetime given by the inverse of the total decay rate, and finally branching ratios. These three quantities are among the properties that distinguish each unstable particle. Therefore, we will see how these quantities are affected by the strong EM field. The parameters that represent and characterize the laser field are the electric field strength $\mathcal{E}_{0}$ and laser frequency $\omega$. In what follows, we will discuss the effect of each of these two parameters on the three quantities mentioned above. We will start, in this order, by first presenting the results of the total decay rate in the presence of the laser, then the lifetime and finally the branching ratios. The total decay rate is obtained by summing the decay rates for all branches. In order to clarify the size of the laser's effect on the total decay rate, we present here a quantity denoted $R_{\text{with/without}}$ and defined as the ratio between the total decay rate in the presence of the laser field and its equivalent in the absence of the laser field; that is
\begin{equation}\label{ratio}
R_{\text{with/without}}=\frac{\Gamma_{W}^{\text{tot}}(\text{with laser})}{\Gamma_{W}^{\text{tot}}(\text{without laser})}.
\end{equation}
In Table~\ref{tab1}, we give some numerical values of this ratio in terms of  the laser field strength $\mathcal{E}_{0}$ and for three different frequencies (CO$_{2}$ laser: $\hbar\omega=0.117~\text{eV}$, Nd:YAG laser: $\hbar\omega=1.17~\text{eV}$ and He:Ne laser: $\hbar\omega=2~\text{eV}$). The number of exchanged photons over which we have summed in the presence of the laser is $-20\leq s\leq +20$, which means the emission and absorption of the same number of photons ($20$ photons). The field strengths used in this research reach up to $\mathcal{E}_{0}=10^{13}~\text{V/cm}$ as a value approximately equivalent to the maximum intensity $(I=10^{22}~\text{W/cm}^{2})$ that has been achieved in powerful laser source technology. Although the theoretical maximum allowable value is $\mathcal{E}_{0}=10^{16}~\text{V/cm}$ (Schwinger limit), we have simply set the field strength at $10^{13}~\text{V/cm}$ in line with the experimentally available intensities in this domain and to avoid instantaneous creation of electron-positron pairs from vacuum \cite{schwinger1,schwinger2,schwinger3}. 
Based on the Table~\ref{tab1}, we can see that the strong laser field strengths have greatly contributed to the reduction of the total decay rate. For the frequency dependence, we observe that the low-frequency laser affects the total decay rate faster than the high-frequency laser. Generally, producing a high-frequency laser effect needs very high field strengths. This behavior of the total decay rate will be fully reflected in the lifetime, as we will see next. After we have given some clarifications about the effect of the laser field on the total decay rate, we will now turn to the study of the changes in $W$-boson lifetime in the presence of the laser field. The order of magnitude of the lifetime of the intermediate bosons ($W~\text{and}~Z$) is about $10^{-25}~\text{sec}$. The lifetime of a particle is given by the inverse of its total decay rate. Therefore, the behavior of the boson lifetime and the change that will arise in the presence of the EM field can be predicted on the basis of the changes in the total decay rate shown in Table~\ref{tab1}. We have seen that the total decay rate decreases with increasing laser field strength, and therefore the probability of decay diminishes in the presence of the strong EM field, which gives the boson the opportunity to live a long time. Originally, the lifetime of the $W$-boson in the absence of a laser field is too long for an elementary particle with a mass greater than $80$ GeV, but too short to be measured directly.
\begin{table}
\caption{\label{tab2}Numerical values of the laser-assisted $W$-boson lifetime as a function of the laser field strength for various numbers of photons exchanged. The frequency of the laser field is $\hbar\omega=1.17~\text{eV}$.}
\begin{ruledtabular}
\begin{tabular}{ccccc} 
 &\multicolumn{4}{c}{$W$-boson lifetime $\tau_{W}$ $[\text{sec}]$}\\  \cline{2-5}
 $\mathcal{E}_{0}$ $[\text{V/cm}]$ & $-20\leq s\leq +20$ & $-80\leq s\leq +80$ & $-140\leq s\leq +140$ & $-200\leq s\leq +200$ \\ \hline
 $10$       & $3.21945\times10^{-25}$      & $3.21945\times10^{-25}$  & $3.21945\times10^{-25}$      & $3.21945\times10^{-25}$      \\
 $10^{2}$   & $3.21945\times10^{-25}$      & $3.21945\times10^{-25}$  & $3.21945\times10^{-25}$      & $3.21945\times10^{-25}$     \\
 $10^{3}$   & $3.21945\times10^{-25}$      & $3.21945\times10^{-25}$   & $3.21945\times10^{-25}$      & $3.21945\times10^{-25}$    \\
 $10^{4}$   & $3.21945\times10^{-25}$      & $3.21945\times10^{-25}$  & $3.21945\times10^{-25}$      & $3.21945\times10^{-25}$     \\
 $10^{5}$   & $3.21952\times10^{-25}$      & $3.21945\times10^{-25}$  & $3.21945\times10^{-25}$      & $3.21945\times10^{-25}$    \\
 $10^{6}$   & $3.69486\times10^{-25}$      & $3.25627\times10^{-25}$  & $3.2269\times10^{-25}$      & $3.21945\times10^{-25}$   \\
 $10^{7}$   & $1.32448\times10^{-24}$     & $5.29587\times10^{-25}$  & $4.15166\times10^{-25}$      & $3.72499\times10^{-25}$  \\
 $10^{8}$   & $1.04563\times10^{-23}$    & $3.0093\times10^{-24}$  & $1.93529\times10^{-24}$      & $1.33097\times10^{-24}$ \\
 $10^{9}$   & $1.05982\times10^{-22}$   & $2.69836\times 10^{-23}$ & $1.53931\times10^{-23}$      & $1.06366\times10^{-23}$ \\
 $10^{10}$  & $1.07067\times10^{-21}$  & $2.71445\times 10^{-22}$ & $1.55225\times10^{-22}$      & $1.08744\times10^{-22}$ \\
 $10^{11}$  & $1.05722\times 10^{-20}$& $2.70465\times 10^{-21}$ & $1.5511\times10^{-21}$      & $1.08738\times10^{-21}$ \\
 $10^{12}$  & $1.05653\times 10^{-19}$& $2.69545\times 10^{-20}$ & $1.54427\times10^{-20}$      & $1.08148\times10^{-20}$ \\
 $10^{13}$  & $1.06593\times 10^{-18}$& $2.70132\times  10^{-19}$ & $1.54657\times10^{-19}$      & $1.0834\times10^{-19}$
\end{tabular}
\end{ruledtabular}
\end{table}
In Table~\ref{tab2}, we include the numerical values of the changes in the lifetime as a function of the field strength with the exchange of a different number of photons for the laser frequency $\hbar\omega=1.17~\text{eV}$. It appears from this table that the lifetime, at low field strengths, remains constant at its value in the absence of the laser (about $3\times 10^{-25}$ sec). As the field strength increases, it is noticeable that the lifetime increases with respect to the number of exchanged photons. The higher the number of exchanged photons, the influence of the laser on the lifetime decreases until it vanishes when we reach a specific number of photons. Due to our limited computing capabilities, we cannot sum over a large number of photons to include this result. The same behavior has been recently revealed for the pion and $Z$-boson lifetimes \cite{mouslih,jakha}. The debate about the laser effect on the total decay rate and lifetime began for the first time in the $1970$s \cite{ritus3,becker} and was revived at the beginning of the $21$st century \cite{muon1,muon2,muon3} to the present day. The effect of the laser field on the lifetime of the muon was studied by Liu \textit{et al.} by embedding the decaying muon in a strong linearly polarized EM field \cite{muon1}. It was found that the muon lifetime decreased and became shorter (specifically from its normal value $2.2\times 10^{-6}~\text{sec}$ to a value of less than  $5\times 10^{-7}~\text{sec}$). In contrast to the results we have obtained, whether here or in previous works \cite{jakha,mouslih}, where we use a circular polarized laser field and find that the lifetime is longer. Thus, it cannot be denied that the polarization of the laser field plays an important role in the contrast of the results. The same can be said about the various changes that can occur in the differential cross section in laser-assisted scattering processes \cite{imane,liu2014,wang2019} with respect to different polarizations of the laser field. However, this difference in lifetime behavior in both cases of polarization can be widely accepted with reference to the quantum Zeno effect \cite{Hindered,zeno} or the anti-Zeno effect \cite{kofman,antizeno}, depending on whether the decay is decelerated or accelerated.
In addition to the laser field strength, we will see if the laser frequency also has an effect on the lifetime. In Fig.~\ref{fig2}, we show the variations in lifetime in terms of field strength and for three different frequencies. Again, as in the case of the total decay rate, we note that the laser effect decreases at higher frequencies.
\begin{figure}[hbtp]
\centering
\includegraphics[scale=0.8]{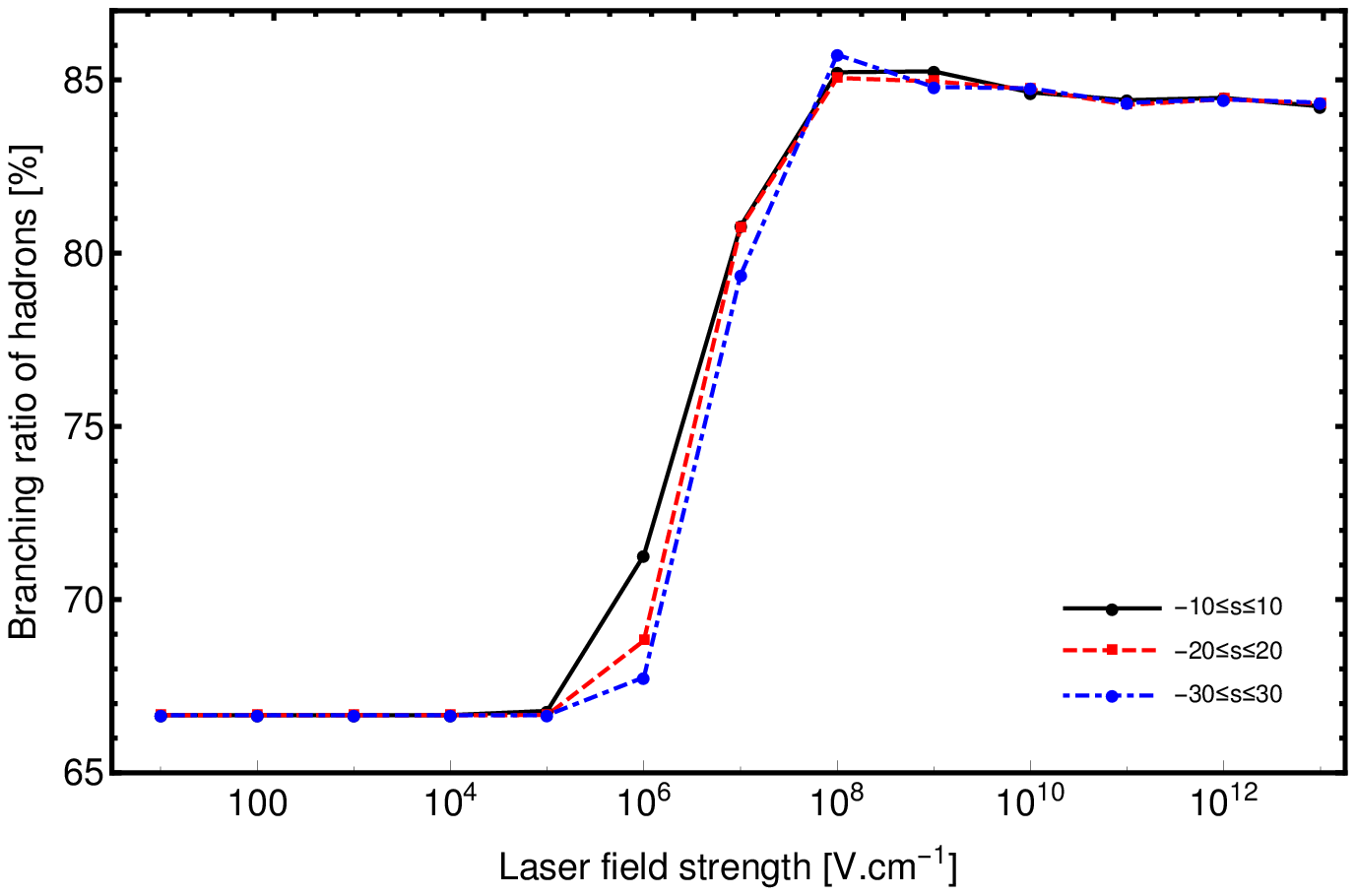}
\caption{The behavior of the branching ratio of hadrons, given in Eq.~(\ref{brhadrons}), as a function of the laser field strength for different numbers of photons exchanged. The frequency of laser field is $\hbar\omega=1.17~\text{eV}$.}\label{fig3}
\end{figure}
\begin{figure}[h!]
\centering
\includegraphics[scale=0.85]{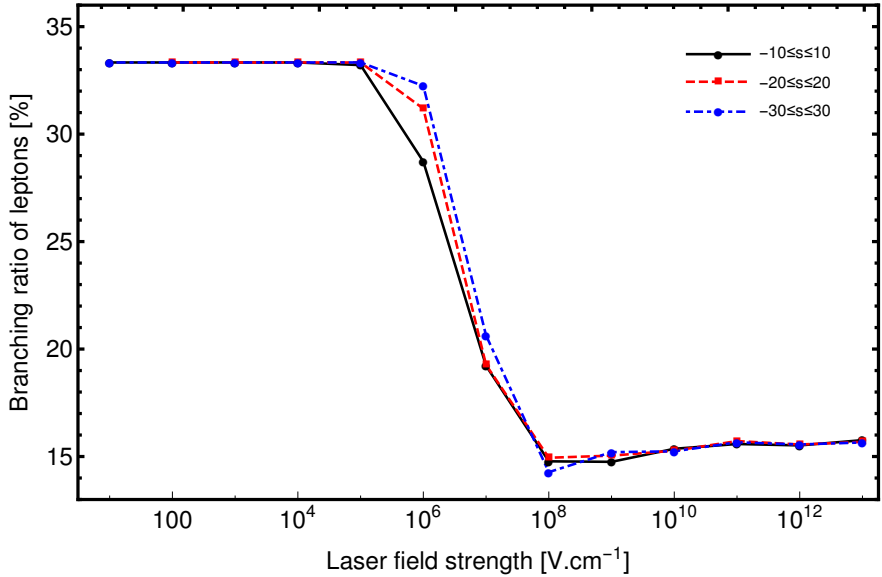}
\caption{The behavior of the branching ratio of leptons, given in Eq.~(\ref{brleptons}), as a function of the laser field strength for different numbers of photons exchanged. The frequency of laser field is $\hbar\omega=1.17~\text{eV}$.}\label{fig4}
\end{figure}
Another important point that also needs to be addressed is how the EM field can influence the BRs and can contribute to their enhancement or suppression. In decay of $W$-boson, we have, as we saw in the previous section, two BRs. One of them is for the hadronic channel and the other for the leptonic one. In the following, we present the effect of laser field strength on each of these two ratios. The BR of hadrons is, in the absence of the laser, the largest between these ratios, meaning that the decay of $W$-boson into a pair of quarks in the absence of an EM field has a high probability compared to the other pairs. The behavior of both $\text{BR}(W^{-}\rightarrow \text{hadrons})$ and $\text{BR}(W^{-}\rightarrow \text{leptons})$ and their variations in terms of laser field strength for three different numbers of exchanged photons are shown, respectively, in Figs.~\ref{fig3} and \ref{fig4}. It can be seen from these two figures that all the corresponding curves for each given number of exchanged photons start from the normal value of each BR in the absence of the laser field ($\sim66\%$ in Fig.~\ref{fig3} and $\sim33\%$ in Fig.~\ref{fig4}), and then stay constant and identical in the range of field strengths between $10^{2}$ and $10^{5}~\text{V/cm}$. Once the latter value has been exceeded, the curves begin to disagree and differ. In another meaning, the BRs were not affected at all by the low-intensity EM field $(10^{2}-10^{5}~\text{V/cm})$, whatever the number of photons exchanged. Outside this range of field strengths $(\mathcal{E}_{0}>10^{5}~\text{V/cm})$, we notice that the BR of hadrons, which is dominant in the absence of the laser, increases and that of leptons decreases until both reach a saturated value ($\sim85\%$ in Fig.~\ref{fig3} and $\sim15\%$ in Fig.~\ref{fig4}) where they are almost stable. This means that the laser contributed to the enhancement of the hadronic BR by making it more dominant, while the leptonic one was weakened and reduced. The increase in the BR of hadrons was balanced by the decrease in that of leptons, which is normal so that the sum of the two ratios remains equal to $100\%$. Therefore, it turns out that the BR of leptons has decreased to the same extent that the BR of hadrons has increased. This significant change in BRs when the EM field is present is a very important and unusual result. Experimental scientists should pay attention to this to ensure and verify its accuracy. Theoretically, the change in BR behavior is mainly due to the difference in phase space of the final particles in the two decay channels. The hadronic decay channel appears to be more stable with the energy reservoir provided by the EM field. 
\begin{table}
\caption{\label{tab3}Numerical values of leptonic (\ref{leptrate}) and hadronic (\ref{hadrate}) decay rates as a function of the laser field strength $\mathcal{E}_{0}$ for a number of exchanged photons $-20\leq s\leq +20$. The frequency of laser field is $\hbar\omega=1.17~\text{eV}$.}
\begin{ruledtabular}
\begin{tabular}{ccc} 
$\mathcal{E}_{0}$ $[\text{V/cm}]$ & $\Gamma(W^{-}\rightarrow \text{leptons})~[\text{GeV}]$  & $\Gamma(W^{-}\rightarrow \text{hadrons})~[\text{GeV}]$  \\ \hline
 $10$       & $0.681497$        &  $1.36299$   \\
 $10^{2}$   & $0.681497$       &   $1.36299$   \\
 $10^{3}$   & $0.681497$       &   $1.36299$   \\
 $10^{4}$   & $0.681497$       &    $1.36299$   \\
 $10^{5}$   & $0.681454$       &   $1.36299$   \\
 $10^{6}$   & $0.555517$       &   $1.22591$  \\
 $10^{7}$   & $0.095728$       & $0.401231$ \\
 $10^{8}$   & $0.00940642$     &  $0.0535423$  
\end{tabular}
\end{ruledtabular}
\end{table}
From Table~\ref{tab3}, we note that both the hadronic and leptonic decay rates decrease with increasing EM field strength, but the hadronic one always remains higher than the leptonic one, which explains the behavior of the BRs shown in Figs.~\ref{fig3} and \ref{fig4}. These findings clearly illustrate how an external EM field could fundamentally change the standard physics of quantum processes in the vacuum and act as a kind of catalyst for new physics and a number of nontrivial phenomena. 
\begin{table}
\caption{\label{tab4}Numerical values of the squared Bessel functions $J_{s}^{2}(z)$ and $J_{s+1}^{2}(z)$ summed over various numbers of exchanged photons for different laser field strengths $\mathcal{E}_{0}$. The frequency of laser field is $\hbar\omega=1.17~\text{eV}$.}
\begin{ruledtabular}
\begin{tabular}{ccccccc} 
&\multicolumn{2}{c}{$-10\leq s\leq +10$}&\multicolumn{2}{c}{$-20\leq s\leq +20$}&\multicolumn{2}{c}{$-30\leq s\leq +30$}\\  \cline{2-7}
$\mathcal{E}_{0}$ $[\text{V/cm}]$ & $J_{s}^{2}(z)$  & $J_{s+1}^{2}(z)$ & $J_{s}^{2}(z)$ & $J_{s+1}^{2}(z)$ & $J_{s}^{2}(z)$ & $J_{s+1}^{2}(z)$ \\ \hline
 $10$       & $1$        &  $1$   & $1$ & $1$ & $1$ &$1$ \\
 $10^{2}$   & $1$       &   $1$   & $1$ & $1$ &$1$ &$1$ \\
 $10^{3}$   & $1$       &   $1$  & $1$& $1$ & $1$  &$1$  \\
 $10^{4}$   & $1$       &    $1$ &$1$ &$1$ & $1$ &  $1$ \\
 $10^{5}$   & $1$       &   $1$   & $1$&$1$  & $1$ &$1$ \\
 $10^{6}$   & $0.999326$ &   $0.997721$  & $1 $ & $1$& $1$ & $1$\\
 $10^{7}$   & $0.0884456$ & $0.0876473$ &$0.176802 $ & $0.170476$ &$0.267065$  & $ 0.258847 $ \\
 $10^{8}$   & $0.00861296$&  $0.00893359$  & $0.0168299 $& $0.0174297$  & $0.0250733$ & $ 0.0259046 $
\end{tabular}
\end{ruledtabular}
\end{table}
For more clarification, the reader may ask why $\mathcal{E}_{0}=10^{6}~\text{V/cm}$ is a critical value delineating no effect versus significant effect seen in previous figures. This can be answered by consulting the numerical values of the summed Bessel functions $J_{s}^{2}(z)$ and $J_{s+1}^{2}(z)$ listed in Table~\ref{tab4}. An examination of the squared amplitude $|\overline{\mathcal{M}^{s}_{fi}}|^{2}$ expression in Eq.~(\ref{result}) reveals that the laser field strength $\mathcal{E}_{0}$ and frequency $\omega$ are involved in determining the decay rate behavior through the argument of Bessel function $z\propto \mathcal{E}_{0}/\omega^{2}$ given in Eq.~(\ref{argument}). A large part of the behavior of the three measurable quantities studied here, such as oscillations and abrupt falls or peaks, is due to the introduction of Bessel functions throughout the theoretical calculation. 
According to Table~\ref{tab4}, it appears that both Bessel functions, summed over the number of photons $s$, take a constant value (e.g., $\sum_{s=-10}^{10}J_{s}^{2}(z)=\sum_{s=-10}^{10}J_{s+1}^{2}(z)=1$) at the interval of low field strengths. Then, as the field strength increases, they begin to change depending on the number of photons exchanged. However, as the photon number increases, the effect of the laser field strength on these functions gradually vanishes. Moreover, it is clearly shown that $10^{6}~\text{V/cm}$ is the threshold value of field strength at which the laser of frequency $\hbar\omega=1.17~\text{eV}$ begins to have an effect. To avoid any confusion, we note here that the values listed in Table~\ref{tab4} do not imply that the properties of the Bessel functions are the only responsible for changes in the quantities studied, but rather are part of other parameters such as the effective mass and the energy-impulsion reservoir provided by the laser field.
\section{Conclusion}\label{sec:conclusion}
In order to highlight the effect of a powerful laser on the total decay rate and lifetime as well as on the branching ratios of $W^{-}$-boson decay, we calculated, recently and in a separate paper, the leptonic decay of the $W^{-}$-boson $(W^{-}\rightarrow \ell^{-} \bar{\nu}_{\ell})$ in the presence of a circularly polarized electromagnetic field \cite{mouslih1}. Then, as a complement to this work, we have focused in the present paper on the laser-assisted hadronic decay of the $W^{-}$-boson $(W^{-}\rightarrow q \bar{q}')$. By combining the results presented here with those obtained previously, we found that the laser considerably extended the lifetime of the boson $W^{-}$ by reducing its total decay rate to a great extent. Yet another interesting result obtained in this study is that the laser has increased the branching ratio of hadrons from
its normal value of $67\%$ to about $85\%$ resulting in a diminution in that of leptons by the same extent. Furthermore, we point out that very rare decay channels that have not been considered in this work are still neglected and have no significant contribution to the total decay rate even in the presence of the laser. These results obtained, particularly related to branching ratios, are very important and require an experimental investigation to support them in order to meet the needs of the scientific community in the future in parallel with the remarkable development of laser technology.
\begin{appendices}
\section*{Appendix: Explicit expression of $|\overline{\mathcal{M}^{s}_{fi}}|^{2}$}\label{appendix}
The detailed and explicit expression obtained for $|\overline{\mathcal{M}^{s}_{fi}}|^{2}$ is given by
\begin{equation}\label{result}
\begin{split}
|\overline{\mathcal{M}^{s}_{fi}}|^{2}=&\frac{1}{3}\Big[AJ_{s}^{2}(z)+BJ_{s+1}^{2}(z)+CJ_{s-1}^{2}(z)+DJ_{s}(z)J_{s+1}(z)+EJ_{s}(z)J_{s-1}(z)\\&+FJ_{s+1}(z)J_{s-1}(z)\Big],
\end{split}
\end{equation}
where the six coefficients $A$, $B$, $C$, $D$, $E$ and $F$ are explicitly expressed by
\begin{equation}
\begin{split}
A=&\dfrac{4}{(k.p)^{2} M_{W}}\big[  (a^{4} (e^{2} - 4 (k.p)^{2})^{2} (k.p_{1})(k.p_{2}) + 2 a^{2} (e^{2} - 4 (k.p)^{2}) (2 (k.p_{1})(k.p_{2})\\ 
&\times M_{W}^{2} + (k.p)^{2} (p_{1}.p_{2}) - (k.p) (k.p_{2})(p.p_{1}) -(k.p) (k.p_{1}) (p.p_{2})) + 2 (k.p)^{2} (M_{W}^{2}\\ 
&\times (p_{1}.p_{2})+ 2 (p.p_{1}) (p.p_{2})))\big],
\end{split}
\end{equation}
\begin{equation}
\begin{split}
B=&\dfrac{-e^2}{(k.p_{1})^2 (k.p_{2})^2 M_{W}^2}\big[2 (k.p_{1})(k.p_{2})\big( (a_{1}.p_{1})(a_{1}.p_{2})(k.p)(-(k.p_{2})\eta + ((k.p_{1})+ 2 (k.p)\eta) \eta')\\
&+a^2 ((k.p_{2})^2 \eta (-(p.p_{1})+M_{W}^2 \eta)-2 (k.p)^2 (p_{1}.p_{2}) \eta \eta'+(k.p_{1})^2 \eta'((p.p_{2})+M_{W}^2\eta')\\
&+(k.p_{1})(k.p_{2})((p.p_{2}) \eta-((p.p_{1})+2 M_{W}^2\eta)\eta')+(k.p)((k.p_{1})\eta'(-(p_{1}.p_{2})+2 (p.p_{2})\eta\\
&+2 (p.p_{1})\eta')+(k.p_{2})\eta((p_{1}.p_{2})+2(p.p_{2})\eta+2(p.p_{1})\eta')))\big)-(k.p_{1})(p_{1}.p_{2})(-(k.p_{2})^2\eta\\
&-(k.p)(k.p_{2})\eta\eta'+(k.p)(k.p_{1})\eta'^2)\epsilon(a_{1},a_{2},k,p) 
+(k.p_{2})((k.p_{2})\eta((k.p_{1})(p.p_{2})+2(k.p_{2})\\
&\times M_{W}^2\eta+4 (k.p)(p.p_{2}) \eta)+2 (k.p_{1})(-2 (k.p_{2})M_{W}^2+(k.p)(p.p_{2}))\eta\eta'-6 (k.p_{1})^2 M_{W}^2 \eta'^2)\\
&\times\epsilon(a_{1},a_{2},k,p_{1})-(k.p_{1})(k.p_{2})((k.p_{2})^2\eta+(k.p_{2}) ((k.p_{1})-5 (k.p)\eta) \eta'-(k.p)(k.p_{1})\eta'^2)\\
&\times\epsilon(a_{1},a_{2},p,p_{1})-(a_{1}.p_{2})(k.p_{1})^2(k.p_{2})\eta' 
\epsilon(a_{2},k,p,p_{1})+(k.p_{1})(p.p_{1})\eta'(2 (k.p_{1})(k.p_{2})\\
&+(k.p)(k.p_{2}) \eta+(k.p)(k.p_{1})\eta')\epsilon(a_{1},a_{2},k,p_{2})+(k.p_{1})^3\eta'(-(k.p_{2})+2 (k.p)\eta')\\
&\times\epsilon(a_{1},a_{2},p,p_{2}) +(a_{1}.p_{1})(k.p_{1})((k.p_{2})^2\eta+(k.p_{2}) ((k.p_{1})+(k.p)\eta)\eta'-(k.p)(k.p_{1})\eta'^2)\\
&\times\epsilon(a_{2},k,p,p_{2})-(k.p)(k.p_{1})(2 (k.p_{2})^2\eta-(k.p_{2})((k.p_{1})+3(k.p)\eta) \eta'+3 (k.p)(k.p_{1})\eta'^2)\\
&\times\epsilon(a_{1},a_{2},p_{1},p_{2})\big],
\end{split}
\end{equation}
\begin{equation}
\begin{split}
C=&\dfrac{e^2}{(k.p_{1})^2 (k.p_{2})^2 M_{W}^2}\big[2 (k.p_{1})(k.p_{2}) \big((a_{1}.p_{1})(a_{1}.p_{2})(k.p)((k.p_{2})\eta-((k.p_{1})+2 (k.p)\eta) \eta') \\
&+a^2 (-(k.p_{2})\eta((k.p)(p_{1}.p_{2})-(k.p_{2})(p.p_{1})(k.p_{1})(p.p_{2})+(k.p_{2}) M_{W}^2 \eta+2(k.p)(p.p_{2})\eta)\\
&+((k.p_{1})((k.p)(p_{1}.p_{2})+(k.p_{2})(p.p_{1})- (k.p_{1})(p.p_{2})+2 ((k.p_{1})(k.p_{2})M_{W}^2+(k.p)^2\\
&\times(p_{1}.p_{2})-(k.p)(k.p_{2})(p.p_{1})-(k.p)(k.p_{1})(p.p_{2}))\eta)\eta'-(k.p_{1})((k.p_{1}) M_{W}^2 + 2 (k.p) \\
&\times(p.p_{1}))\eta'^2)\big)-(k.p_{1})(p_{1}.p_{2})(-(k.p_{2})^2\eta-(k.p) (k.p_{2})\eta\eta' +(k.p)(k.p_{1})\eta'^2)\\
&\times\epsilon(a_{1},a_{2},k,p) +(k.p_{2})((k.p_{2})\eta((k.p_{1})(p.p_{2})+2(k.p_{2}) M_{W}^2\eta+4(k.p)(p.p_{2})\eta)+2 (k.p_{1})\\
&\times(-2(k.p_{2}) M_{W}^2 + (k.p)(p.p_{2})) \eta \eta'-6 (k.p_{1})^2 M_{W}^2\eta'^2)\epsilon(a_{1},a_{2},k,p_{1})-(k.p_{1})(k.p_{2})\\
&\times((k.p_{2})^2\eta+(k.p_{2})((k.p_{1})-5(k.p)\eta)\eta'-(k.p)(k.p_{1})\eta'^2)\epsilon(a_{1},a_{2},p,p_{1})-(a_{1}.p_{2})\\
&\times(k.p_{1})^2(k.p_{2})\eta'\epsilon(a_{2},k,p,p_{1})+(k.p_{1})(p.p_{1})\eta'(2 (k.p_{1}) (k.p_{2})+(k.p)(k.p_{2})\eta+(k.p)\\
&\times(k.p_{1})\eta'\epsilon(a_{1},a_{2},k,p_{2})
+(k.p_{1})^3\eta'(-(k.p_{2})+2(k.p)\eta')\epsilon(a_{1},a_{2},p,p_{2})+(a_{1}.p_{1}) (k.p_{1})\\
&\times((k.p_{2})^2\eta+(k.p_{2})((k.p_{1})+(k.p)\eta)\eta'-(k.p)(k.p_{1}) \eta'^2)\epsilon(a_{2},k,p,p_{2}) -(k.p)(k.p_{1})\\
&\times(2 (k.p_{2})^2 \eta-(k.p_{2}) ((k.p_{1})+3(k.p)\eta) \eta'+3 (k.p)(k.p_{1}) \eta'^2)\epsilon(a_{1},a_{2},p_{1},p_{2}) \big],
\end{split}
\end{equation}
\begin{equation}
\begin{split}
D=&\dfrac{-2 e }{(k.p)(k.p_{1}) (k.p_{2}) M_{W}^2}\big[a^2 e^2 (k.p_{1})(k.p_{2})((a_{1}.p_{2})((k.p_{1})+2(k.p)\eta) +(a_{1}.p_{1})((k.p_{2})\\
&-2(k.p)\eta')) -2 (a^2 (e^2 - 4 (k.p)^2)(k.p_{2})-2(k.p)(p.p_{2})) ((k.p_{2})\eta+(k.p_{1})\eta')\epsilon(a_{2},k,p,p_{1})\\
&+(-a^2 (e^2 -4 (k.p)^2)(k.p_{2})((k.p_{1})+2(k.p)\eta)-2 (k.p)M_{W}^2((k.p_{2})\eta-3 (k.p_{1})\eta'))\\
&\times\epsilon(a_{2},k,p_{1},p_{2})\big],
\end{split}
\end{equation}
\begin{equation}
\begin{split}
E=&\dfrac{2 e}{(k.p)(k.p_{1})(k.p_{2})M_{W}^2}\big[2 (a_{1}.p_{1})(k.p)(k.p_{2}) ((k.p_{1})(p.p_{2})+(k.p_{2})M_{W}^2\eta+2(k.p)(p.p_{2})\eta \\
&\times (k.p_{1}) M_{W}^2 \eta')-2 (a_{1}.p_{2})(k.p)(k.p_{1})(-(k.p_{2})(p.p_{1})+ (k.p_{2})M_{W}^2 \eta+(k.p_{1}) M_{W}^2\eta'\\
&+2 (k.p)(p.p_{1})\eta')-a^2 (e^2 - 4 (k.p)^2)(k.p_{1})(k.p_{2})((a_{1}.p_{2})((k.p_{1})+2(k.p)\eta)+(a_{1}.p_{1})\\
&\times((k.p_{2}) -2(k.p)\eta'))-2 (a^2 (e^2 - 4 (k.p)^2)(k.p_{2})-2 (k.p)(p.p_{2}))((k.p_{2})\eta+(k.p_{1})\eta')\\
&\times\epsilon(a_{2},k,p,p_{1})
+ (-a^2 (e^2 -4 (k.p)^2)(k.p_{2})((k.p_{1})+2(k.p)\eta)-2 (k.p) M_{W}^2 ((k.p_{2})\eta\\
&-3 (k.p_{1})\eta')\epsilon(a_{2},k,p_{1},p_{2})+2 (k.p)(k.p_{1})(k.p_{2})\epsilon(a_{2},p,p_{1},p_{2})\big],
\end{split}
\end{equation}
\begin{equation}
\begin{split}
F=-\dfrac{4 (a_{1}.p_{1})(a_{1}.p_{2}) e^2(k.p)(-(k.p_{2})\eta+((k.p_{1})+2(k.p)\eta) \eta')}{(k.p_{1})(k.p_{2}) M_{W}^2},
\end{split}
\end{equation}
where the different scalar products are evaluated in the rest frame of $W^{-}$-boson; and for all 4-vectors $a, b, c$ and $d$, we have
\begin{equation}
\epsilon(a,b,c,d)=\epsilon^{\mu\nu\rho\sigma}a_{\mu}b_{\nu}c_{\rho}d_{\sigma},
\end{equation}
where $\epsilon^{\mu\nu\rho\sigma}$ is the antisymmetric tensor with the convention $\epsilon^{0123}=1$. The reader may refer to our previous works \cite{mouslih,jakha} to see how these tensors are calculated analytically.
\end{appendices}

\end{document}